\documentclass[conference]{IEEEtran}
\IEEEoverridecommandlockouts
\usepackage{enumitem}
\usepackage{cite}
\usepackage{amsmath,graphicx,amsfonts}
\usepackage{algorithmic}
\usepackage{enumitem}
\usepackage{graphicx}
\usepackage{textcomp}
\usepackage{xcolor}
\usepackage{csquotes}
\usepackage{balance}
\usepackage{url}
\usepackage{balance}

\def\BibTeX{{\rm B\kern-.05em{\sc i\kern-.025em b}\kern-.08em
    T\kern-.1667em\lower.7ex\hbox{E}\kern-.125emX}}

\usepackage{soul, color}
\usepackage[acronym,toc,shortcuts]{glossaries}
\usepackage{multirow}

\usepackage{caption}
\usepackage{subcaption}

\usepackage{booktabs}

\begin{document}

\title{Kolmogorov-Arnold Networks (KANs) \\ for Time Series Analysis}

\author{Cristian J. Vaca-Rubio, Luis Blanco, Roberto Pereira, Màrius Caus \\
{\normalsize{} Centre Tecnològic de Telecomunicacions de Catalunya (CTTC/CERCA), Castelldefels, Barcelona, Spain, 08860.} \\
{\normalsize{} Emails: \texttt{\{cvaca, lblanco, rpereira, mcaus\}@cttc.es}}

\thanks{This work was funded by the European Commission under the “5G-STARDUST” Project, which received funding from the Smart Networks and Services Joint Undertaking (SNS JU) under the European Union’s Horizon Europe research and innovation programme under Grant Agreement No. 101096573, and in part by the grant CHIST-ERA-20-SICT-004 (SONATA) by PCI2021-122043-2A/AEI/10.13039/501100011033. This work has been submitted to IEEE for possible publication. Copy
right may be transferred without notice, after which this version may no
 longer be accessible.
}}

\maketitle

\begin{abstract}
This paper introduces a novel application of Kolmogorov-Arnold Networks (KANs) to time series forecasting, leveraging their adaptive activation functions for enhanced predictive modeling. Inspired by the Kolmogorov-Arnold representation
theorem, KANs replace traditional linear weights with spline-parametrized univariate functions, allowing them to learn activation patterns dynamically. 
We demonstrate that KANs outperforms conventional Multi-Layer Perceptrons (MLPs) in a real-world satellite traffic forecasting task, providing more accurate results with considerably  fewer number of learnable parameters. We also provide an ablation study of KAN-specific parameters impact on performance. The proposed approach opens new avenues for adaptive forecasting models, emphasizing the potential of KANs as a powerful tool in predictive analytics.

\end{abstract}
\section{Introduction}
\label{sec:intro}
Time series forecasting is a traditional problem that plays a key role in a wide range of fields, driving critical decision-making processes in finance, economics, medicine, meteorology, and biology, reflecting the wide applicability and its significance across many domains \cite{sezer2020financial, prakarsha2022time, chen2023long, zhu2023weather2k}. It involves predicting future values based on the previously observed data points. With this goal in mind, understanding the dynamics of time-dependent phenomena is essential and requires unveiling the patterns, trends and dependencies hidden with the historical data. While conventional approaches have been traditionally centered on parametric models grounded in domain-specific knowledge, such as autoregressive (AR), exponential smoothing, or structural time series models, contemporary Machine Learning (ML) techniques offered a pathway to discern temporal patterns solely from data-driven insights.

Non-ML methods traditionally tackle the time series forecasting problem and often rely on statistical methods to predict future values based on previously observed data.
One of the most well-known techniques is the AutoRegressive Integrated Moving Average (ARIMA) model, which combines auto-regression, integration, and moving averages to forecast data. The authors in \cite{box2015time} detailed this approach, providing a comprehensive methodology foundational for subsequent statistical forecasting methods. Extensions of ARIMA, like Seasonal ARIMA (SARIMA), adapt the model to handle seasonality in data series, particularly useful in fields like retail and climatology \cite{hyndman2018forecasting}. Exponential Smoothing techniques constitute another popular set of traditional (non-ML-based) forecasting methods. They are characterized by their simplicity and effectiveness in handling data with trends and seasonality. An exponent of this family of techniques is the so-called Holt-Winters seasonal technique, which adjusts the model parameters in response to changes in trend and seasonality within the time series data \cite{holt2004forecasting, winters1960forecasting}. 
These models have been widely used for their efficiency,  interpretability
and implementation.

More recently, ML models have significantly impacted the forecasting landscape by handling large datasets and capturing complex nonlinear relationships that traditional methods cannot. In recent years, Deep Learning (DL)-based forecasting models \cite{lim2021time, torres2021deep} have gained popularity, motivated by the notable achievements in many fields.  For instance, neural networks have been extensively studied due to their flexibility and adaptability. Simple Multi-Layer Perceptron (MLPs) were among the first to be applied to forecasting problems, demonstrating significant potential in non-linear data modeling \cite{zhang2012neural, chen2023long}.

Built upon these light models, more complex architectures have progressively expanded the capabilities of neural networks in time series forecasting. Typical examples are  recurrent neural network architectures such as  Long Short-Term Memory (LSTM) networks and Gated Recurrent Units (GRUs), which are designed to maintain information in memory for long periods without the risk of vanishing gradients -- a common issue in traditional recurrent networks \cite{hochreiter1998vanishing, hochreiter1997long}. On a related note,  Convolutional Neural Networks (CNNs), which are fundamentally inspired by MLPs, are also extensively employed in time series forecasting. These architectures are particularly efficient at processing temporal sequences due to their strong spatial pattern recognition capabilities.
The combination of CNNs with LSTMs has resulted in models that efficiently process both spatial and temporal dependencies, enhancing forecasting accuracy \cite{borovykh2017conditional}. These models have started to outperform established benchmarks in complex forecasting tasks, motivating a significant shift towards more complex network structures. Unfortunately, as the majority of the models mentioned above are inspired by MLP architecture, they tend to have poor scaling law \cite{bachmann2024scaling}, i.e., the number of parameters in MLPs networks do not scale linear with the number of  layers, and often  lack interpretability.

A recent study in reference \cite{liu2024kan}, which caught the attention of the research community, introduces Kolmogorov-Arnold Networks (KANs), a novel neural network architecture designed to potentially replace traditional multilayer perceptrons. 
KANs represent a disruptive paradigm shift, and as a potential game changer have recently attracted the interest of the AI community worldwide.  They are inspired by the Kolmogorov-Arnold representation theorem \cite{kolmogorov1961representation,braun2009constructive, schmidt2021kolmogorov}. Unlike MLPs, which are inspired by the universal approximation theorem, KANs take advantage of this representation theorem to generate a different architecture. They innovate by replacing linear weights with spline-based univariate functions along the edges of the network, which are structured as learnable activation functions. This design not only enhances the accuracy and interpretability of the networks, but also enables them to achieve comparable or superior results with smaller network sizes across various tasks, such as data fitting and solving partial differential equations. 
While  KANs show promise in improving the efficiency and interpretability of neural network architectures, 
the study acknowledges the necessity for further research into their robustness when applied to diverse datasets and their compatibility with other deep learning architectures. These areas are crucial for understanding the full potential and limitations of KANs.

Our paper is a prospective study that investigates the application of KANs to time series forecasting. We aim to evaluate the practicality of KANs in real-world scenarios, to the best of the authors' knowledge, not previously explored in the literature, analyzing their efficiency regarding the number of trainable parameters and discussing how the additional degrees of freedom might affect forecasting performance. Herein, we will assess the performance using real-world satellite traffic data. This exploration seeks to further validate KANs as a versatile tool in advanced neural network design for time series forecasting, although more comprehensive studies are required to optimize their use across broader applications.
Finally, we note that due to the early stage of KANs, it is fair to compare them as a potential alternative to MLPs, but further investigation is needed to develop more complex solutions that can compete with advanced architectures such as LSTMs, GRUs, and CNNs, already well-established on the MLP-based architectures \cite{livieris2020cnn, mehtab2022analysis}.

This paper is structured as follows. Section 2 presents the problem statement, providing fundamental background on the Kolmogorov-Arnold representation theorem and describes our generalized KANs for time series forecasting. Section 3 introduces the experimental setup description. Simulation results analyzing the performance of KANs with real-world datasets are shown in Section 4. Finally, concluding remarks are provided in Section 5. 

\section{Problem statement} 
\begin{figure}
     \centering
        \includegraphics[width=\linewidth]{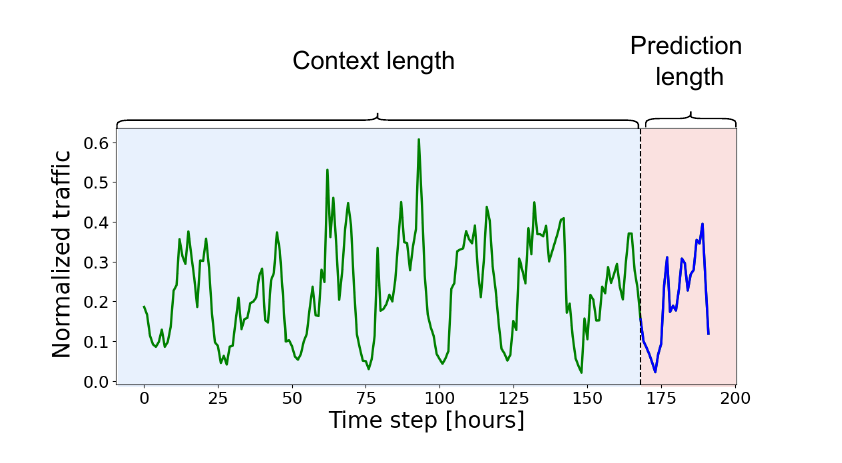}
\caption{Example of  normalized satellite traffic series data with the conditioning and prediction lengths denoted in blue, and red, respectively.}
\label{fig:time_series_example}
\end{figure}

We formulate the traffic forecasting problem as a time series at time $t$ represented by $y_{t}$. Our objective is to predict the future values of the series
\begin{equation}
\mathbf{y}_{t_0:T} = [y_{t_0}, y_{t_0+1}, ..., y_{t_0+T}] \label{eq:pred_range}
\end{equation}
based solely on its historical values
\begin{equation}
\mathbf{x}_{t_0-c:t_0-1} = [x_{t_0-c}, ..., x_{t_0-2}, x_{t_0-1}] \label{eq:cond_range}
\end{equation}
where $t_0$ denotes the starting point from which future values $y_{t}, t=t_0,..., T$ are to be predicted. We differentiate the historical time range $[t_0-c, t_0-1]$ and the forecast range $[t_0, T]$ as the context and prediction lengths, respectively. Our approach focuses on generating point forecasts for each time step in the prediction length, aiming to achieve accurate and reliable forecasts. Figure \ref{fig:time_series_example} shows an exemplary time series.
\subsection{Kolmogorov-Arnold representation background} \label{sec:KAT}
Contrary to MLPs, which are based on universal approximation theorem, KANs rely on the Kolmogorov-Arnold representation theorem, also known as the Kolmogorov-Arnold superposition theorem. A fundamental result in the theory of dynamical systems and ergodic theory. It was independently formulated by Andrey Kolmogorov and Vladimir Arnold in the mid-20th century. 

The theorem states that any multivariate continuous function $f$, which depends on $\mathbf{x}=[x_1, x_2,…, x_n]$, on a bounded domain, can be represented as the finite composition of simpler continuous functions, involving only one variable. Formally, a real, smooth, and continuous multivariate function $f(\mathbf{x}):[0,1]^{n}\rightarrow \mathbb{R}$ can be represented by the finite superposition of univariate functions \cite{kolmogorov1961representation}: 
\begin{equation}
    f(\mathbf{x}) =\sum_{i=1}^{2n+1} \Phi_i\left(\sum_{j=1}^n\phi_{i,j}(x_j)\right), \label{eq:K-ART}
\end{equation}
where $\Phi_i:\mathbb{R}\to\mathbb{R}$ and $\phi_{i,j}:[0,1]\to\mathbb{R}$ denote the so-called outer and inner functions, respectively. One might initially perceive this development as highly advantageous for ML. The task of learning a high-dimensional function simplifies to learning a polynomial number of one dimensional functions.  Nevertheless, these 1-dimensional functions can exhibit non-smooth characteristics, rendering them potentially unlearnable in practical contexts. As a result of this problematic behavior, the Kolmogorov-Arnold representation theorem has been traditionally disregarded in machine learning circles, recognized as theoretically solid, but ineffective in practice. Unexpectedly, the theoretical result in \cite{liu2024kan} has recently emerged as a potential game changer, paving the way for new network architectures, inspired by the Kolmogorov-Arnold theorem. 
\subsection{Kolmogorov-Arnold network background}
The authors in \cite{liu2024kan} mention that equation (\ref{eq:K-ART}) has two layers of non-linearities, with $2n+1$ terms in the middle layer. Thus, we only need to find the proper functions inner univariate functions $\phi_{i,j}$ and $\Phi_i$ that approximate the function. The one dimensional inner functions $\phi_{i,j}$ can be approximated using B-splines. A spline is a smooth curve defined by a set of control points or knots. Splines are often used to interpolate or approximate data points in a smooth and continuous manner. A spline is defined by the order $k$ ($k=3$ is a common value), which refers to the degree of the polynomial functions used to interpolate or approximate the curve between control points. The number of intervals, denoted by $G$, refers to the number of segments or subintervals between adjacent control points. In spline interpolation, the data points are connected by these segments to form a smooth curve (of $G+1$ grid points). Although splines other than B-splines could also be considered, this is the approach proposed in \cite{liu2024kan}. Equation (\ref{eq:K-ART}) can be represented as a 2-layer (or analogous 2-depth) network, with activation functions placed at the edges (instead of at the nodes) and nodes performing a simple summation. Such two-layer network is too simplistic to effectively approximate any arbitrary function with smooth splines. For this reason, reference 
\cite{liu2024kan} extends the ideas discussed above by 
proposing a generalized 
architecture with wider and deeper KANs.

\begin{table*}[]
\centering
\caption{Model configurations for satellite traffic forecasting}
\resizebox{\textwidth}{!}{%
\begin{tabular}{@{}ccccc@{}}
\toprule
\textbf{Model} & \textbf{Configuration}    & \textbf{Time horizon (h)}       & \textbf{Spline details}                & \textbf{Activations} \\ \midrule
MLP (3-depth)  & [168, 300, 300, 300, 24] & Context/Prediction: 168/24 & N/A                                    & ReLU (fixed)                \\
MLP (4-depth)  & [168, 300, 300, 300, 300, 24] & Context/Prediction: 168/24 & N/A                                    & ReLU (fixed)                 \\
KAN (3-depth)  & [168, 40, 40, 24]      & Context/Prediction: 168/24 & Type: B-spline, $k = 3$, $G=5$ & learnable                  \\
KAN (4-depth)  & [168, 40, 40, 40, 24] & Context/Prediction: 168/24 & Type: B-spline, $k = 3$, $G=5$ & learnable                  \\
\bottomrule
\end{tabular}
}
\label{tab:params}
\end{table*}

A KAN layer is defined by a matrix $\mathbf{\Phi}$ \cite{liu2024kan} composed by univariate functions $\{\phi_{i,j} (\cdot)\}$ with $i=1,...,N_{in}$ and $j=1,...,N_{out}$, where $N_{in}$ and $N_{out}$ denote the number of inputs and the number of outputs, respectively, and $\phi_{i,j}$ are the trainable spline functions described above. Note according to the previous definition, the Kolmogorov-Arnold representation theorem presented in Section \ref{sec:KAT} can be expressed as a two-layer KAN.  The inner functions constitute a KAN layer with $N_{in}=n$ and $N_{out}=2n+1$, while the external functions constitute another KAN layer with $N_{in}=2n+1$ and $N_{out}=1$.

Let us define the shape of a KAN by $[n_1,...,n_{L+1}]$, where $L$ denotes the number of layers of the KAN. It is worth noting the Kolmogorov-Arnold theorem is defined by a KAN of shape $[n,2n+1,1]$. A generic deeper KAN can be expressed by the composition $L$ layers: 
\begin{equation}
    \mathbf{y} = \text{KAN}(\mathbf{x}) = (\mathbf{\Phi}_L \circ \mathbf{\Phi}_{L-1} \circ \ldots \circ \mathbf{\Phi}_{1}) \mathbf{x}.
\end{equation}
Notice that all the operations are differentiable. Consequently, KANs can be trained with backpropagation. Despite their elegant mathematical foundation, KANs are simply combinations of splines and MLPs, which effectively exploit each other's strengths while mitigating their respective weaknesses. Splines stand out for their accuracy on low-dimensional functions and allow transition between various resolutions. Nevertheless, they suffer from a major dimensionality problem due to their inability to effectively exploit compositional structures. In contrast, MLPs experience a lower dimensionality problem, due to their ability to learn features, but exhibit lower accuracy than splines in low dimensions due to their inability to optimize univariate functions effectively. KANs have by their construction 2 levels of degrees of freedom. Consequently, KANs possess the capability not only to acquire features, owing to their external resemblance to MLPs, but also to optimize these acquired features with a high degree of accuracy, facilitated by their internal resemblance to splines. To learn features accurately, KANs can capture compositional structure (external degrees of freedom), but also effectively approximate univariate functions (internal degrees of freedom with the splines). It should be noted that by increasing the number of layers $L$ or the dimension of the grid $G$, we are increasing the number of parameters and, consequently, the complexity of the network.  This approach constitutes an alternative to traditional DL models, which are currently relying on MLP architectures and motivates our extension of this work.

\subsection{KAN time series forecasting network}
We frame our traffic forecasting problem as a supervised learning framework consisting of a training dataset with input-output $\{\mathbf{x}_{t_0-c:t_0-1}, \mathbf{y}_{t_0:T}\}$ in the condition and prediction lengths. We want to find $f$ that approximates $\mathbf{y}_{t_0:T}$, i.e., $\mathbf{y}_{t_0:T} \approx f (\mathbf{x}_{t_0-c:t_0-1})$. 
For ease of notation, we describe our framework as a two-layer (2-depth) KAN [$N_{i}$, $n$, $N_{o}$](note that to comply with the original paper notation, the input layer is not accounted as a layer per se). The output and input layers will be comprised of $N_{o}$, and $N_{i}$ nodes corresponding to the total amount of time steps in \eqref{eq:pred_range} and \eqref{eq:cond_range}, while the transformation/hidden layer of $n$ nodes.  The inner functions constitute a KAN layer with $N_{in}=N_i$ and $N_{out}=n$, while the external functions constitute another KAN layer with $N_{in}=n$ and $N_{out}=N_o$. Our KAN can be expressed by the composition of 2 layers: 
\begin{equation}
    \mathbf{y} = \text{KAN}(\mathbf{x}) = (  \mathbf{\Phi}_{2} \circ  \mathbf{\Phi}_{1}) \mathbf{x},
\end{equation}
where the output functions $\Phi_2$ generates the $N_o$ outputs values corresponding to \eqref{eq:pred_range} by doing the transformation from the previous layers, i.e, we predict the $T$ time steps. The proposed network can be used to forecast future traffic data in the prediction length, based solely on the context length. 

\begin{figure}[t]
         \centering
        \includegraphics[width=\linewidth]{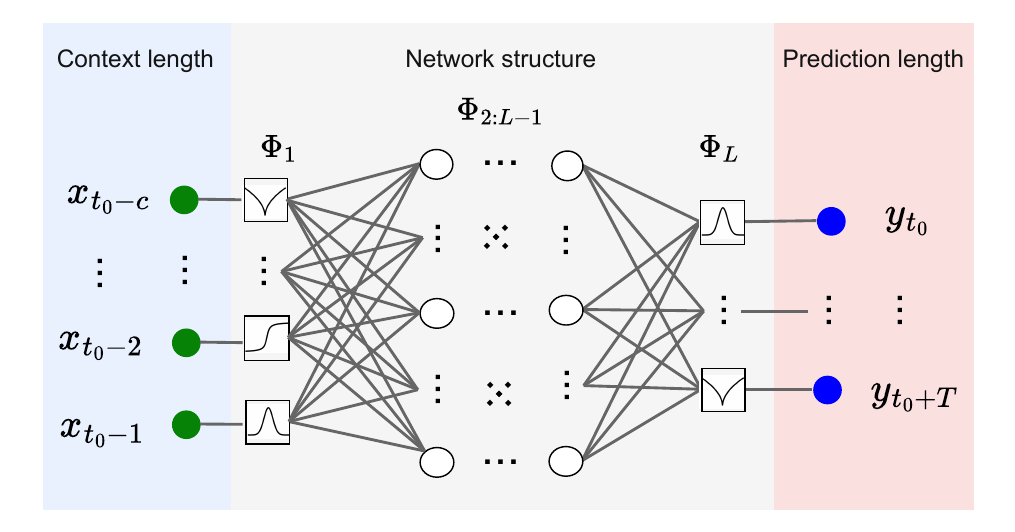}
\caption{Example of the flow of information in the KAN network architecture for our traffic forecasting task. Learnable activations are represented inside a square box.
}
\label{fig:kans_architecture}
\end{figure}

Fig. \ref{fig:kans_architecture} shows a generic representation for any arbitrary number of layers $L$ as presented in (4). 

\section{Experimental setup}
The dataset has been generated within the context of the European project 5G-STARDUST. The inputs are obtained from a satellite operator (SO), as a result of processing real information from a GEO satellite communication system, which provisions broadband services. The dataset is a long time series capturing aggregated traffic data. To preserve privacy, anonymous clients have been defined with more than 500 connected users, and the traffic has been normalized. The measurements are monthly long, and the time granularity is 1 hour. The traffic has been extracted per satellite beam in Megabits per second (Mbps).  Although the data has been collected using a GEO satellite communication system, it is expected that user needs could be used to address LEO systems, as well. It is worth emphasizing that the data collected can be used for AI-driven predictive analysis, to forecast traffic conditions, which is essential to avoid congestion and to make efficient use of satellite resources. Endowing the network with intelligence will be beneficial to meet the different demands of satellite applications. 

We aim to investigate the forecasting performance of different KAN and MLP architectures for predicting satellite traffic over a total of six beam areas. Concretely, we have a context length of 168 hours (one week) and a prediction length of 24 hours (one day). This translates to $T=24$, $c=168$, where $y_{t_0+T} = 192$ in \eqref{eq:pred_range} and \eqref{eq:cond_range}. Our focus is on evaluating the efficacy of KAN models compared to traditional MLPs\footnote{As KANs are in their infancy, we remark this comparison is fair instead of comparing against more complex architectures as LSTM.}. We designed our experiments to compare models with similar depths but varying architectures to analyze their impact on forecasting accuracy and parameter efficiency. Table \ref{tab:params} summarizes the parameters selected for this evaluation. We have data for the six beams over one month. We use two weeks + 1 day for training and one week + 1 day for testing for all the different beams on the dataset. These test series were not seen by the network during training time. We train all the networks with $500$ epochs and Adam optimizer with a learning rate of $0.001$. The selected loss function minimizes the mean absolute error (MAE) of the values around the prediction length.

\begin{figure}[h!]
     \centering
     \begin{subfigure}[b]{0.5\textwidth}
        \includegraphics[width=\linewidth]{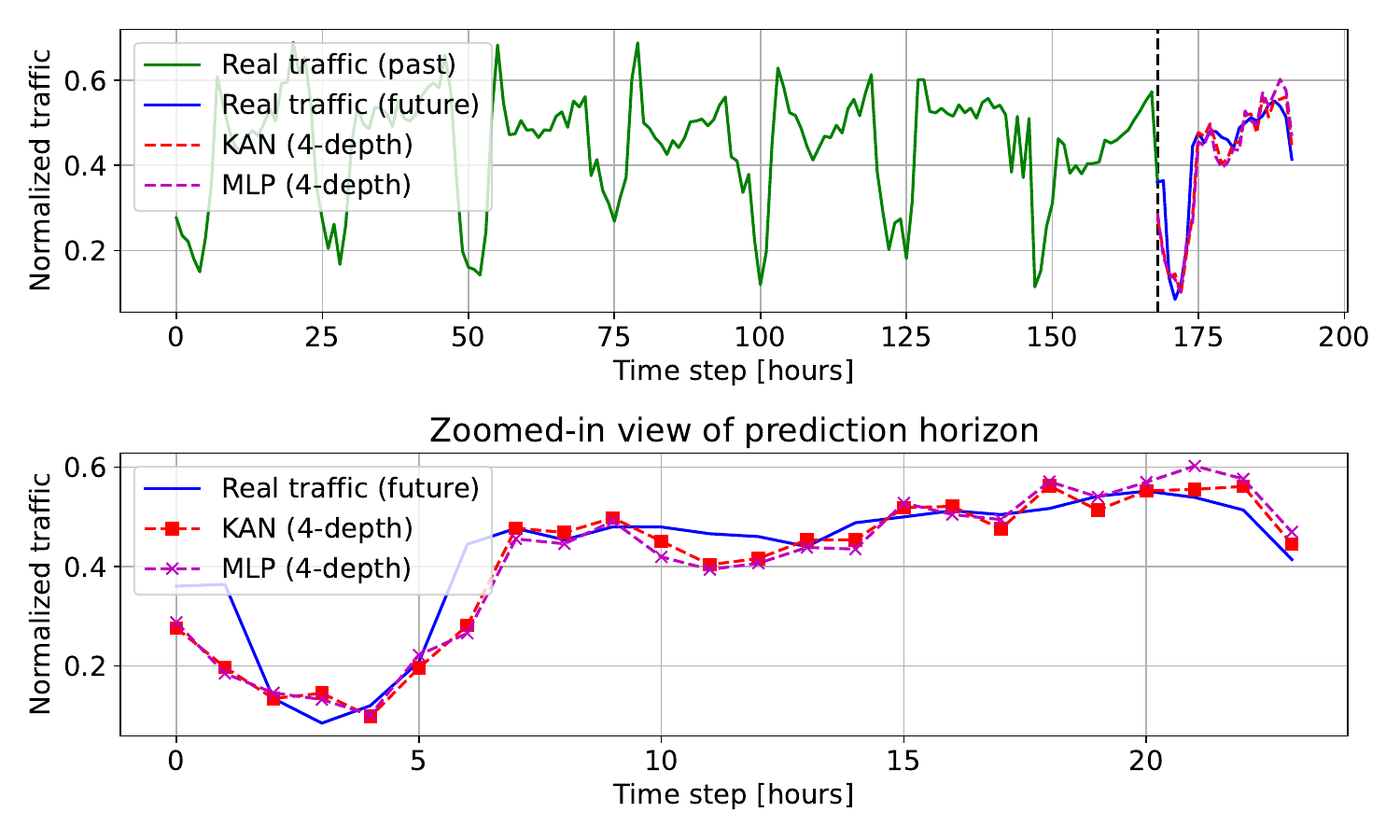}
    \caption{Forecast over beam 1.}
    \label{fig:f_Plot_1}
     \end{subfigure}
     \begin{subfigure}[b]{0.5\textwidth}
         \centering
        \includegraphics[width=\linewidth]{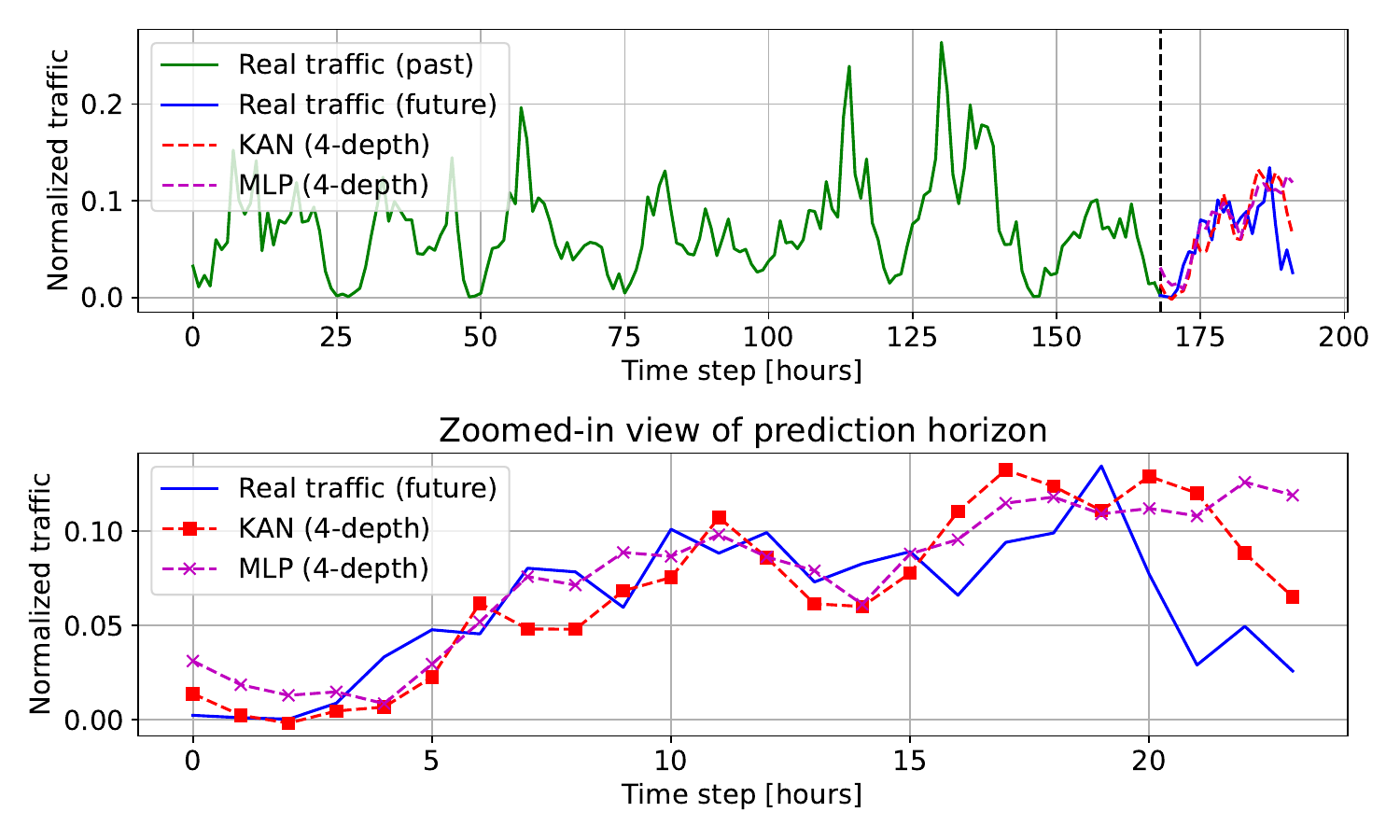}
        \caption{Forecast over beam 2.}
        \label{fig:f_Plot_2}
     \end{subfigure}
     \begin{subfigure}[b]{0.5\textwidth}
         \centering
        \includegraphics[width=\linewidth]{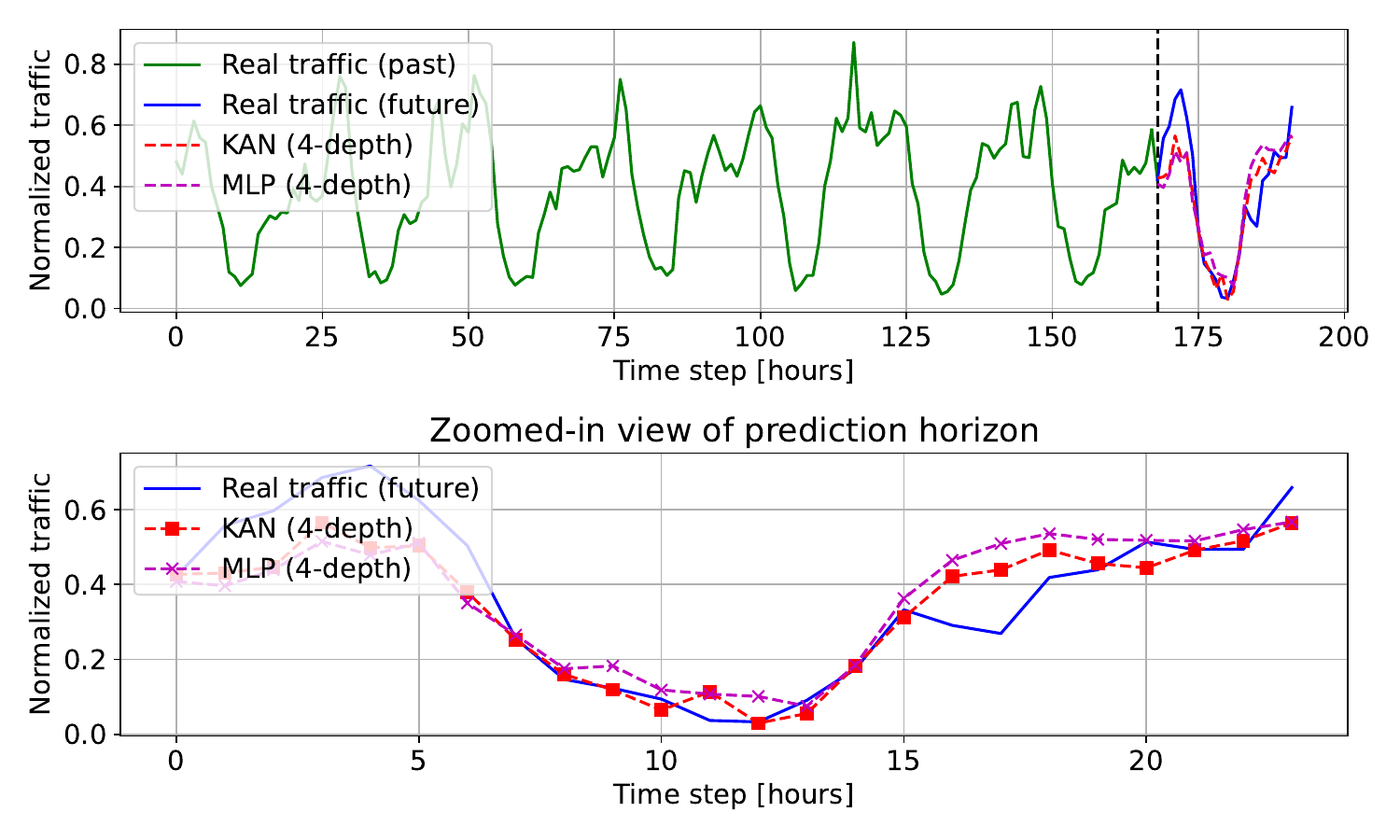}
        \caption{Forecast over beam 3.}
        \label{fig:f_Plot_3}
     \end{subfigure}
\caption{Satellite traffic over three different beams with their forecasted values using a 4-depth KAN and a 4-depth MLP.}
\label{fig:forecast_plots}
\end{figure}

\section{Simulation results}

\subsection{Performance analysis}
We analyze the forecasting performance in the prediction length for different beams over the test set. Figures 
\ref{fig:forecast_plots}a-c
depicts the real traffic value used as input (in green) to the networks, the expected output prediction length (in blue) and the values predicted values using a  KAN (in red) and MLP (in purple) of depth 4 both -- see Table~\ref{tab:params} for details on model configuration. In general, our results show that the predictions obtained using KANs better approximates the real traffic values than the predictions obtained using traditional MLPs.

This is particularly evident in Figure \ref{fig:f_Plot_1}. Here, KAN accurately matches rapid changes in traffic volume, which the MLP models sometimes moderately over/under-predicted, as the last part of the forecast shows. This capability suggests that KANs are better suited to adapt to sudden shifts in traffic conditions, a critical aspect of effective traffic management. 

Additionally, the responsiveness of KANs is particularly noticeable in Figure \ref{fig:f_Plot_2} during fast changing traffic conditions. KAN shows a rapid adjustment to its forecast that is closely aligned with the actual traffic pattern. This is particularly noticeable in the last $6$ hours of the prediction length where MLP exhibits a lag failing to capture these immediate fluctuations, which shows its worse 
performance to capture 
dynamic traffic variations.
Further analysis is shown in Figure \ref{fig:f_Plot_3}, where traffic conditions are more variable and intense, demonstrated the robustness of KAN in maintaining high performance despite the complexity and higher volume. This robustness suggests that KANs can manage different scales and intensities of traffic data more effectively than MLPs, making them more reliable for deployment in varied traffic scenarios.

To further quantify the performance and advantages of using KANs for the satellite traffic forecasting task we show Table \ref{tab:model_outputs}. It shows a detailed comparison of MLPs and KANs different architectures used for evaluation over all the beams. The table displays the Mean Squared Error (MSE), Root Mean Squared Error (RMSE), Mean Absolute Error (MAE), Mean Absolute Percentage Error (MAPE), and the number of trainable parameters for each model. Analyzing the error metrics, it becomes  clear that KANs outperform MLPs, where the KAN (4-depth) is the best in performance. Its lower values in MSE and RMSE indicates its better ability to predict traffic volumes with lower deviation. Similarly, its lower values in MAE and MAPE suggests that KANs not only provides more accurate predictions but also maintains consistency across different traffic volumes, which is crucial for practical traffic forecasting scenarios.

Furthermore, the parameter count reveals a significant difference in model complexity. KAN models are notably more parameter-efficient, with KAN (4-depth) utilizing only 109k parameters compared to 329k parameters for MLP (4-depth) or 238k for MLP (3-depth). \textbf{This reduced complexity suggests that KANs can achieve higher or comparable forecasting accuracy with simpler and potentially lighter models}. Such efficiency is especially valuable in scenarios where computational resources are limited or where rapid model deployment is required. The results also show that with an augmentation of 16k parameters in KAN, the performance can be significantly improved, contrary to MLPs which an increment of 91k parameters does not showcase a significant improvement.

From a technical perspective, KANs leverage a theoretical foundation that provides an intrinsic advantage in modeling complex, non-linear patterns typical in traffic systems. This capability likely contributes to their flexibility and accuracy in traffic forecasting. The consistency in performance across diverse conditions also suggests that KANs have strong generalization capabilities, which is essential for models used in geographically varied locations under different traffic conditions. Moreover, besides obtaining lower error rates, our results also suggest that KANs can do so with considerably smaller number of parameters than traditional MLP networks.

\begin{table}[]
\caption{Results summary}
\centering
\resizebox{\columnwidth}{!}{%
\begin{tabular}{@{}cccccc@{}}
\toprule
\textbf{Model} & \textbf{MSE} ($\times10^{-3}$) & \textbf{RMSE} ($\times10^{-2}$) & \textbf{MAE} ($\times10^{-2}$) & \textbf{MAPE} & \textbf{Parameters} \\\midrule
MLP (3-depth) & 6.34 & 7.96 & 5.41 & 0.64 & 238k \\
MLP (4-depth) & 6.12 & 7.82 & 5.55 & 1.05 & 329k \\
KAN (3-depth) & 5.99 & 7.73 & 5.51 & 0.62 & 
\textbf{93k} \\
KAN (4-depth) & \textbf{5.08} & \textbf{7.12} & \textbf{5.06} & \textbf{0.52} & 109k \\
      \bottomrule
      \label{tab:model_outputs}
\end{tabular}
}
\end{table}
\subsection{KANs parameter-specific analysis}
We provide an insightful analysis of how different configurations of nodes and grid sizes affect the performance of KANs, particularly in the context of traffic forecasting. For this analysis, we designed 3 KANs (2-depth) $[168, n, 24]$ with $n \in \{5, 10, 20\}$ and varying grids $G \in \{5, 10, 20\}$ for a $k=3$ order B-spline. These results are shown during training time.

\begin{figure}[t!]
\centering
\includegraphics[width=\linewidth]{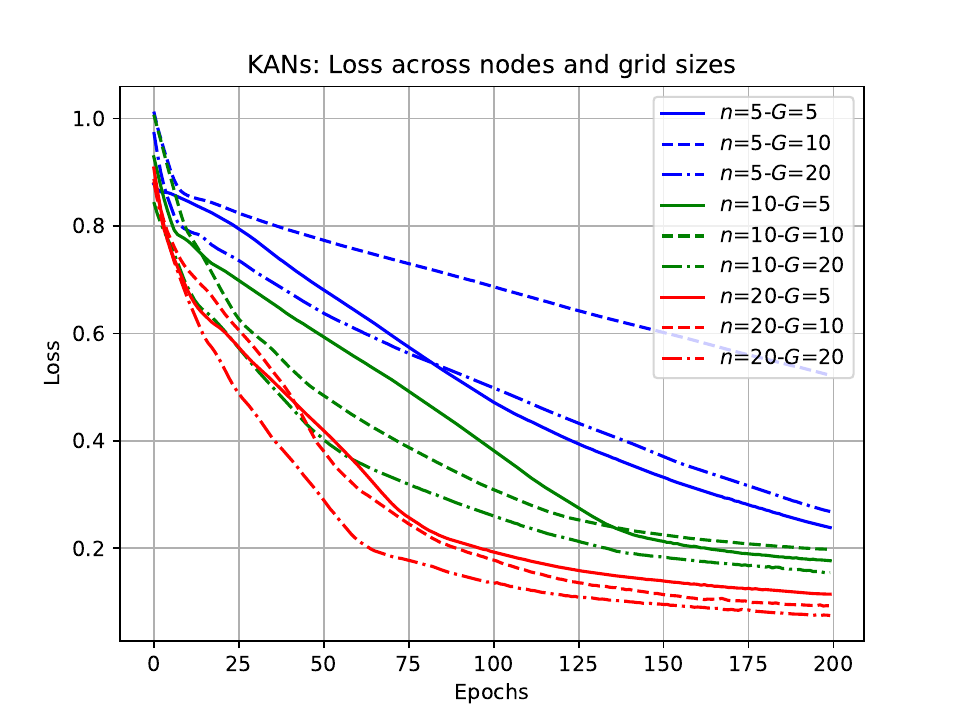}
\caption{Ablation comparison of KAN-specific parameters \\ during training time.}
\label{fig:f_Plot_4}
\end{figure}

Figure \ref{fig:f_Plot_4} shows a clear trend where increasing the number of nodes generally results in lower loss values. This indicates that higher node counts are more effective at capturing the complex patterns in traffic data, thus improving the performance. For instance, configurations with $n=20$ demonstrate significantly lower losses across all grid sizes compared to those with fewer nodes.

Similarly, the \textbf{grid size within the splines of KANs has a notable impact on model performance}. Larger grid sizes, when used with a significant amount of nodes ($n \in \{10, 20\}$), consistently result in better performance. However, when the amount of nodes is low ($n=5$) the extra complexity of the grid size shows the opposite effect. When having a significant amount of nodes larger grids likely provide a more detailed basis for the spline functions, allowing the model to accommodate better variations in the data, which is crucial for capturing complex temporal traffic patterns.

The best performance is observed in configurations that combine a high node count with a large grid size, such as the $n=20$, and $G=20$ setup. This combination likely offers the highest degree of flexibility and learning capacity, making it particularly effective for modeling the intricate dependencies found in traffic data. However, this superior performance comes at the cost of potentially higher computational demands and longer training times, as more trainable parameters are included.

These findings imply that while increasing nodes and grid sizes can significantly enhance the performance of KANs, these benefits must be weighed against the increased computational requirements. For practical applications, particularly in real-time traffic management where timely responses are critical, it is essential to strike a balance. An effective approach could involve starting with moderate settings and gradually adjusting the nodes and grid sizes based on performance assessments and computational constraints. Besides, we want to highlight that for this study continual learning was not assessed, a possibility mentioned in the original paper \cite{liu2024kan}. 

\section{Conclusion}
In this paper, we have performed an analysis of KANs and MLPs for satellite traffic forecasting. The results highlighted several benefits of KANs, including superior forecasting performance and greater parameter efficiency. In our analysis, we showed that KANs consistently outperformed MLPs in terms of lower error metrics and were able to achieve better results with lower computational resources. Additionally, we explored specific KAN parameters impact on performance. This study showcases the importance of optimizing node counts and grid sizes to enhance model performance. Given their effectiveness and efficiency, KANs appear to be a reasonable alternative to traditional MLPs in traffic management. 
\bibliographystyle{ieeetr}
\bibliography{refs}  

\end{document}